\documentclass[final]{iopart}
\usepackage{graphicx}
\usepackage{subfig}
\usepackage{iopams}

\newcommand{\ket}[1]{|#1 \rangle}
\newcommand{\bra}[1]{\langle #1|}

\begin{document}
\title{Coherent Tunneling Adiabatic Passage with the
Alternating Coupling Scheme}

\author{L M Jong, A D Greentree, V I Conrad, L C L Hollenberg and D N Jamieson}
\address{Centre for Quantum Computer Technology, School of Physics,
  University of Melbourne, VIC 3010, Australia}
\ead{lmjong@unimelb.edu.au}
\date{\today}
\begin{abstract}
The use of adiabatic passage techniques to mediate particle transport through real space, rather than phase space, is becoming an interesting possibility. We have investigated the properties of Coherent Tunneling Adiabatic
Passage (CTAP) with alternating tunneling matrix elements.  This
coupling scheme, not previously considered in the donor in silicon paradigm,
provides an interesting route to long-range quantum transport. We
introduce simplified coupling protocols and transient eigenspectra as
well as a realistic gate design for this transport
protocol. Using a pairwise treatment of the tunnel couplings for a 5 donor
device with $30~\textrm{nm}$ donor spacings, 120nm total chain length, we
estimate the time scale required for adiabatic operation to be $\sim70~\textrm{ns}$,
a time well within measured electron spin and estimated charge
relaxation times for phosphorus donors in silicon. 
 
\end{abstract}


\maketitle

\section{Introduction}

The potential of silicon based proposals in particular to build upon
and integrate with existing CMOS fabrication techniques makes them an
attractive candidate for large scale quantum information processing
architectures. Since the original proposal by Kane 
\cite{bib:KaneNature1998} for qubits based on the nuclear spins of
$^{31}\mathrm{P}$ donors in isotopically pure silicon
$^{28}\mathrm{Si}$, other donor based systems have been suggested
based on electron charge \cite{bib:HollenbergPRB2004} and spin
\cite{bib:VrijenPRA2000, bib:HillPRB2005, bib:deSousaPRA2004} degrees of freedom. Scale-up,
however, requires more than the ability to fabricate many qubits and
the associated nanoelectronics. In Ref. \cite{bib:HollenbergPRB2006} a
scalable donor electron spin qubit architecture was proposed based on
a bi-linear arrangement of qubits. A key feature of this architecture
is a transport mechanism using Coherent Tunneling Adiabatic Passage
(CTAP) \cite{bib:GreentreePRB2004} which both allows for qubit
transport during quantum error correction and serves to reduce the
effective gate density of the nanoelectronics. CTAP is a spatial
analogue of the well known STIRAP protocol \cite{bib:VitanovARPC2001}
from quantum optics for coherent transfer of quantum information. The advantage of CTAP is that it affords long-range, flexible and robust transfer as is required for scale up of donor based architectures.

In addition to phosphorus in silicon, there are proposals to observe
CTAP in quantum dots
\cite{bib:PetrosyanOptComm2006,bib:ShroerPRB2007}, superconductors
\cite{bib:SiewertOptComm2006}, single atoms in optical potentials
\cite{bib:EckertPRA2004, bib:EckertOptCom2006}, and Bose-Einstein
Condensates \cite{bib:GraefePRA2006,bib:RabPRA2008}.  Recently Longhi et
al. \cite{bib:LonghiJPhysB2007,bib:LonghiPRB2007,bib:DellaValleAPL2008} demonstrated CTAP of photons in a
three-waveguide structure adding significant motivation to
demonstrations with massive particles. CTAP has also been proposed
as a means of implementing a quantum analogue to fanout using MRAP (Multi-Recipient Adiabatic Passage) \cite{bib:GreentreePRA2006,bib:DevittQIP2007} and operator
  measurements. Other recent works have investigated the differences
  in CTAP in quantum optics and solid state systems
  \cite{bib:ColePRB2008,bib:Opatrny2008}. Atomistic simulations of
  triple donor systems in silicon have been investigated \cite{bib:Rahman2009}, verifying the existence
  of an CTAP pathway for a small three donor device and the use of
  ion-implantation for fabrication of CTAP devices have been discussed
  Ref. \cite{bib:VanDonkelaar2008}. Triple dots have recently also
  been demonstrated in GaAs 2DEG structures
  \cite{bib:GaudreauPRL2006,bib:RoggePRB2008,bib:AmahaAPL2009} and carbon nanotubes
  \cite{bib:GroveNanoLett2008}.

  The transport protocols discussed in
  \cite{bib:GreentreePRB2004,bib:HollenbergPRB2006} were based on the
  straddling scheme introduced by Malinovsky and Tannor
  \cite{bib:MalinovskyPRA1997} where two sites are coupled to the ends
  of a strongly coupled and uncontrolled chain.  Providing that the
  tunneling matrix elements between sites on the chain are
  significantly higher than the gate controlled tunneling matrix
  elements (TMEs) of the end-of-chain sites, then the protocol works by
  transporting the particle between the end-of-chain sites with
  negligible population ever appearing along the intermediate sites in
  the chain (even transiently).

 In this paper we consider a different adiabatic protocol suitable for long range
 (multi-site) transport based on the alternating scheme. This was proposed
 in the context of quantum dot transport of electrons by Petrosyan and Lambropolous
  \cite{bib:PetrosyanOptComm2006} and is also known for STIRAP protocols \cite{bib:ShorePRA1991} Here we consider this scheme for
  the case of phosphorus ions in silicon.

Successful coherent transfer of the electron across the chain requires the
protocol to be completed within the charge relaxation time, and in order
to be useful for spin based quantum computing architectures must also
be within the qubit spin dephasing time, $T_2$. Recent
measurements of electron spin relaxation for phosphorus donors in purified
$\mathrm{Si^{28}}$ have yielded a $T_2$ of approximately 60ms at 7K
\cite{bib:TyryshkinPRB2003} and still in the millisecond range when
the donors are placed near a surface or interface \cite{bib:TyryshkinPhE2006} . Charge dephasing in a Si:P-P$^+$ system
have been investigated in Refs. \cite{bib:BarrettPRB2003,
  bib:GormanPRL2005}, estimating the dephasing time to be of order nanoseconds.

\begin{figure}
\begin{centering}
\includegraphics[height=3cm]{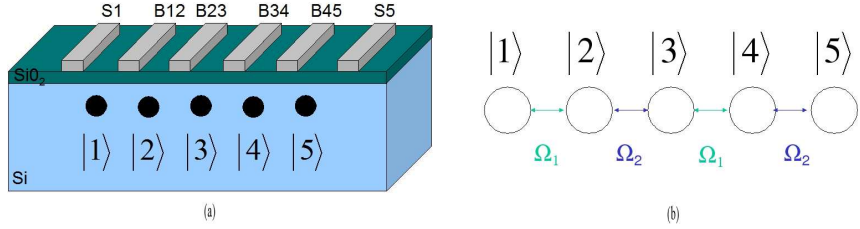}
\caption{(a) Schematic of a five donor, one electron device for CTAP using
  the alternating coupling scheme. Barrier gates ($\mathrm{B_{ij}}$) control the
  tunneling of the electron between buried phosphorus donor sites. (b)
  The counter intuitive coupling scheme. Tunneling matrix elements in the five site chain are controlled
as an A-B chain. The counter-intuitive pulse sequence applies all of the $\Omega_2$
tunneling before $\Omega_1$ tunneling to effect transport from $\ket{1}$ to $\ket{5}$.
\label{fig:schematic}
}
\end{centering}
\end{figure}

We consider a five donor structure, illustrated in Figure
 \ref{fig:schematic}(a), where four donors are ionized and one is neutral. Gates on the surface control the tunnel matrix
 elements and thereby define the adiabatic pathway for electronic
 transport. The counter-intuitive, alternating gate bias sequence,
 illustrated in figure \ref{fig:schematic}(b), varies the tunnel matrix
 elements resulting in robust transport of the electron from one end
 of the chain to the other. This structure is extendable
 to an arbitrary (odd) number of sites. Based on an effective mass
 treatment we analyse the system and make estimates of the
 timescale required for the operation of CTAP with the alternating
 coupling scheme for this device. 

\section{Analytics}
\label{sec:analytics}
We begin by writing down the analytical form of the null space for
the problem with $2n+1$ states for a single particle.  We assume
that the TME between states is $\ket{2i-1}$
and $\ket{2i}$ is $\Omega_i$, and that all of the even $\Omega_i$
and the odd $\Omega_i$ can be controlled separately from 0 to some
maximum value. Note that we are here explicitly using the modal
approximation. Recent calculations have shown this to
be a very good approximation for CTAP problems
\cite{bib:RabPRA2008,bib:ColePRB2008,bib:Opatrny2008,bib:Rahman2009} although the
exact nulling of some central state populations that is predicted for CTAP does \emph{not} precisely occur for realistic potentials.
Assuming that the energies of the states are all degenerate, we write down
the Hamiltonian as:
\begin{equation}
\mathcal{H}=\sum_{i=1}^n \Omega_{2i-1}\ket{2i-1}\bra{2i} +
\Omega_{2i}\ket{2i}\bra{2i+1} + \mathrm{h.c.} 
\label{eq:hamiltonian}
\end{equation}
We can immediately write down the null space solution, which we write
as $\ket{\mathcal{D}_0}=\sum_{i=1}^{2n+1}\psi_i\ket{i}$.  First note
that all of the $\psi$ at even sites must vanish, i.e. $\psi_{2i} = 0$.
Secondly we can write down the values of the $\psi$ at odd sites as a series,
i.e.
\begin{eqnarray}
\psi_{2i+1} = - \frac{\Omega_{2i-1}}{\Omega_{2i}} \psi_{2i-1},
\end{eqnarray}
and hence
\begin{eqnarray}
\psi_{2i+1} = (-1)^{i} \prod_{j = 1}^{j = i}
\left(\frac{\Omega_{2j-1}}{\Omega_{2j}}\right) \psi_1 .
\end{eqnarray}
We may now write down the normalisation to get the correct form of
the null vector, $\ket{\mathcal{D}_0}$, which is
\begin{eqnarray}
\ket{\mathcal{D}_0} &=& \frac{1}{N}\left(\prod_{i=1}^n
\Omega_{2i}\ket{1} + \cdots +(-1)^j
\prod_{i=j}^{n}\Omega_{2i}\prod_{i=1}^{j}\Omega_{2i-1}\ket{2j+1}\right
. \nonumber \\
&+& \cdots + \left . (-1)^{n}\prod_{i=1}^n \Omega_{2i-1}\ket{2n+1}\right)
\end{eqnarray}

with
\begin{eqnarray}
N &=& \left[ \left(\prod_{i=1}^n \Omega_{2i}\right)^2 + \cdots + \left(\prod_{i=j}^{n}\Omega_{2i}\prod_{i=1}^{j}\Omega_{2i-1} \right)^2\right .\nonumber \\
&+& \cdots + \left . \left(\prod_{i=1}^n \Omega_{2i-1}\right)^2 \right]^{1/2}
\end{eqnarray}
This form for $\ket{\mathcal{D}_0}$ immediately shows that this
null space has all the desired properties.  It is unidimensional,
when the odd TMEs are all zero and the even TMEs all
non-zero, the system is in $\ket{1}$, and when the even TMEs
are all zero with the odd non-zero, the system is in state
$|2n+1\rangle$, with a smooth adiabatic pathway between the
states. Apart from enforcing the zeros of the tunneling matrix element
no control as the \emph{relative} values is required.
In fact, this adiabatic pathway is maintained provided that only the
\emph{ends} of the alternating pathway can be satisfactorily nulled,
ie that $\Omega_1=0$ at t=0 and $\Omega_{2n}=0$ at
$\mathrm{t=t_{\max}}$, which makes the protocol an attractive prospect
for implementing as less fine controls are required, decreasing the
amount of gates and connections required.

Considering the  $\ket{\mathcal{D}_0}$ for a five site system (ACTAP$_5$) explicitly we have: 
\begin{equation}
\ket{\mathcal{D}_0^{(5)}} = \frac{\Omega_2\Omega_4 \ket{1} - \Omega_1 \Omega_4 \ket{3} + \Omega_1 \Omega_3 \ket{5}}{\sqrt{(\Omega_2\Omega_4)^2 + (\Omega_1\Omega_4)^2 + (\Omega_1\Omega_3)^2}}.
\label{eq:dnaught}
\end{equation}
 We see that when $\Omega_1$ is zero, there is only
 population in $\ket{\mathcal{D}_0}=\ket{1}$.  Similarly, when
$\Omega_4$ is zero, the null state is $\ket{\mathcal{D}_0}=\ket{5}$ does not depend on $\Omega_2$.  This allows us to consider global controls of all the even tunneling matrix
elements together and similarly for all the odd elements, with only
the need to fine tune the ends of the chain, $\Omega_1$ and
$\Omega_4$ in the ACTAP$_5$ case. Global controls, and specifically
A-B chains as we are discussing here, have been investigated in the
context of Heisenberg chains \cite{bib:BenjaminPRL2002}.

Another important feature of equation \ref{eq:dnaught} is that it illustrates the robustness of the ACTAP protocol to fabrication errors.  In particular, one should note the existence of the null state irrespective of the values of the TMEs, providing that they can be varied between zero and finite values.  This is useful when considering the limitations of conventional fabrication techniques, e.g. single ion implantation \cite{bib:JamiesonAPL2005} which will have unavoidable variability in the final positions of the donors due to straggle.  As only the ratios of the TMEs enter into the form of the null state, the adiabatic pathway persists in the presence of such variations in site to site coupling.  It is important to note that the adiabaticity of the protocol, and hence the time taken for high-fidelity transport, will vary also, and so any given device will need to be calibrated.  Further discussions on the role of TME variability in related CTAP protocols can be found in \cite{bib:HollenbergPRB2006, bib:Rahman2009, bib:VanDonkelaar2008}.

\begin{figure}
\begin{centering}
\includegraphics[height=10cm]{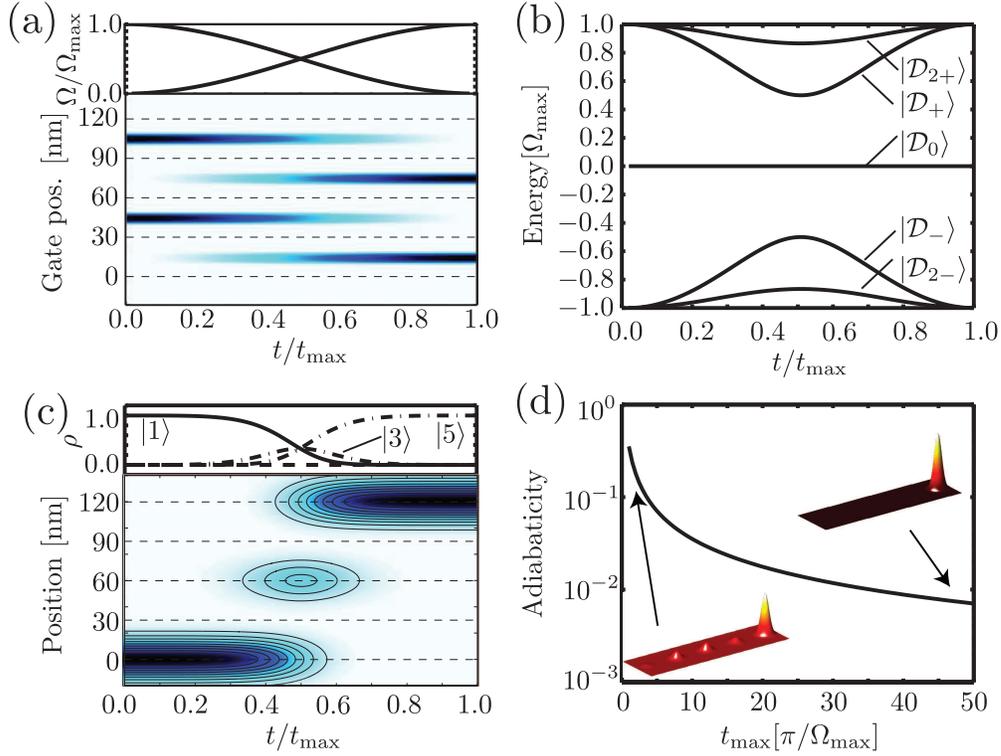}
\caption{(a) The top trace shows the TMEs $\Omega_1$ and $\Omega_2$ as a function of time. Below, the TMEs between donors (postions marked by dashed lines) are illustrated as a density plot as a function of time. TMEs controlled by gates 2 and 4 are high initially ($\Omega_2$), whilst the TMEs controlled by gates 1 and 3 are low initially ($\Omega_1$). Varying the biases on the gates effects the counter-intuitive pulse sequence described. (b) Eigenvalue spectrum of the ACTAP$_5$ Hamiltonian . (c) Populations of
the position eigenstates throughout the protocol. Note the complete
transfer from $\ket{1}$ to $\ket{5}$ but with transient
population in $\ket{3}$, and no population in either $\ket{2}$ or
$\ket{4}$. (d) Adiabaticity as $t_{\max}$ increases. As $t_{\max}$ is
increased the adiabaticity parameter is decreased indicating better
fidelity tranfer across the chain. A shorter total time for the
protocol can to transitions out of the $\ket{\mathcal{D}_0}$ state and
the transfer of the electron from $\ket{1}$ to $\ket{5}$ is no longer complete.
\label{fig:actap5plots}}
\end{centering}
\end{figure}

For simplicity let us now consider the case where
all the odd tunnel matrix elements are the same, and the even ones
also, i.e.
\begin{eqnarray}
\Omega_{2i-1} = \Omega_1 \mbox{ and }  \Omega_{2i} = \Omega_2 \mbox{ }\forall i,
\end{eqnarray}
and concentrating on ACTAP$_5$ we have the following result for the null
state:
\begin{equation}
 \ket{\mathcal{D}_0^{(5)}} = \frac{\Omega_2^2 \ket{1} -
\Omega_1\Omega_2 \ket{3} + \Omega_1^2 \ket{5}}{\sqrt{\Omega_1^4 +
\Omega_2^4 + \Omega_1^2\Omega_2^2}}.
\end{equation}
Furthermore we can also determine all of the
eigenstates and eigenvalues for this symmetric case.  The
eigenstates are
\begin{eqnarray}
\ket{\mathcal{D}_0^{(5)}} &= \frac{\Omega_2^2 \ket{1} - \Omega_1 \Omega_2 \ket{3} + \Omega_1^2\ket{5}}{\sqrt{\Omega_1^4 + \Omega_2^4+ \Omega_1^2\Omega_2^2}},\\
 \ket{\mathcal{D}_{\pm}^{(5)}} &= [-\Omega_1 \ket{1} \mp
  \sqrt{\Omega_1^2-\Omega_1 \Omega_2 + \Omega_2^2} (\ket{2} - \ket{4})
  \nonumber \\
+ &(\Omega_1 - \Omega_2) \ket{3} + \Omega_2\ket{5}][2\sqrt{\Omega_1^2 - \Omega_1\Omega_2 + \Omega_2^2}]^{-1},\\
 \ket{\mathcal{D}_{2\pm}^{(5)}} &= [\Omega_1 \ket{1} \pm \sqrt{\Omega_1^2+\Omega_1 \Omega_2 + \Omega_2^2} (\ket{2} + \ket{4})\nonumber \\
+ &(\Omega_1 + \Omega_2) \ket{3} + \Omega_2\ket{5}][2\sqrt{\Omega_1^2 + \Omega_1\Omega_2 + \Omega_2^2}]^{-1},
\end{eqnarray}

with eigenenergies
\begin{eqnarray}
E_0 &=& 0, \\
E_{\pm} &=& \pm \sqrt{\Omega_1^2 - \Omega_1\Omega_2 + \Omega_2^2}, \\
E_{2\pm} &=& \pm \sqrt{\Omega_1^2 + \Omega_1\Omega_2 + \Omega_2^2}.
\end{eqnarray}
To explore the evolution of the system, for analytical simplicity we
chose square sinusoidal pulse variation of $\Omega_1$ and $\Omega_2$, however in keeping with most adiabatic protocols, the exact form of the variation in tunneling matrix elements is not essential.  For pulses from $t=0$ to $t=t_{\max}$ we set:

\begin{eqnarray}
\Omega_1 & = & \Omega_{\max}\sin^2\left(\frac{\pi t}{2 t_{\max}}\right),\label{eq:tmes}\\
\Omega_2 & = & \Omega_{\max}\cos^2\left(\frac{\pi t}{2 t_{\max}}\right),
\label{eq:tmes2}
\end{eqnarray}
and these pulses are illustrated in Figure \ref{fig:actap5plots}(a).

The null state is the eigenstate which has zero energy throughout the
protocol.  This is the state used for the CTAP protocol, as its
evolution under the pulse sequence gives rise to a smooth change in
population from $\ket{1}$ at $t=0$ to $\ket{5}$ at $t = t_{\max}$. We
can then find the eigenvalues for this Hamiltonian, shown plotted in
figure \ref{fig:actap5plots}(b). Solving the master equation for this
Hamiltonian, we obtain the occupancies of each of the donor sites
throughout the protocol. Analogously with other CTAP protocols, we
treat the electron as initially occupying the first site, and transferred
entirely to the end-of-chain site at the end of the protocol. In the
alternating scheme there is transient occupation of the site in the
middle of the chain, as shown in Figure \ref{fig:actap5plots}(c)  but other intermediate sites remain unoccupied.  This differs from the straddling scheme \cite{bib:GreentreePRB2004,bib:HollenbergPRB2006} where occupation of all intermediate sites is strongly suppressed. 

The Adiabaticity parameter can be fairly easily calculated
between the null state and either of the nearest neighbours, so in
particular, choosing the closest positive state, we write the
Adiabaticity parameter as
\begin{eqnarray}
\mathcal{A} &=& \frac{|\bra{\mathcal{D}_{+}} \frac{\partial
\mathcal{H}}{\partial t} \ket{\mathcal{D}_0}|}{|E_{+}-E_0|^2}, \\
 &=& \frac{\dot{\Omega}_1\Omega_2^2 - \dot{\Omega}_2\Omega_1^2 + \Omega_1\Omega_2(\dot{\Omega}_1 - \dot{\Omega}_2)}{2\sqrt{\Omega_1^4 + \Omega_1^2 \Omega_2^2 + \Omega_2^4}(\Omega_1^2 - \Omega_1\Omega_2 + \Omega_2^2)}
\end{eqnarray}

For adiabatic evolution we require that $\mathcal{A}\ll 1$ with the maximal
adiabaticty determining the time allowable for a realistic experiment. If we note that the adiabaticity parameter will be largest at the
crossing point of the two TMEs, and if we apply
pulses that are symmetric, then we can make a considerable
simplification to the above.  So at the point where $\mathcal{A}$ is
greatest, we have $\Omega_1 = \Omega_2 = \Omega_{\max}/2$, and
$\dot{\Omega}_1 = \dot{\Omega}_2 = \pi\Omega_{\max}/(2t_{\max})$ and
we find using the form of the TMEs as given in equation \ref{eq:tmes}, the above equation gives rise to the simple form

\begin{equation}
\mathcal{A}=\frac{4\pi}{\sqrt{3}\Omega_{\max}t_{\max}}.
\label{eq:adiabatsimpagain}
\end{equation}
This dependence of the adiabaticity parameter on the the total protocol time
$t_{\max}$is plotted in Figure \ref{fig:actap5plots}(d). With
longer $t_{\max}$, and hence lower $\mathcal{A}$, the
transported electron is more likely to remain in the desired
$\ket{\mathcal{D}_0}$ state resulting in better fidelity transfer.

A realistic device may not have enough fine control over the
TMEs to be able to completely turn them on or off. In this
case we have non-zero minimum values and assuming a form for the TMEs of:
\begin{eqnarray}
\Omega_1(t) & = & \Omega_{1\min} + \left( \Omega_{1\max}-\Omega_{1\min}\right)\sin^2\left(\frac{\pi t}{2 t_{\max}}\right),\\
\Omega_2(t) & = & \Omega_{2\min} + \left( \Omega_{2\max}-\Omega_{2\min}\right)\cos^2\left(\frac{\pi t}{2 t_{\max}}\right).
\end{eqnarray}
In the case where complete suppression of the tunneling is not
possible, a measure of our ability to turn off tunneling between donors is the
contrast ratio between the overlap of the initial and final states
with $\ket{\mathcal{D}_0}$. The contrast ratio defines an error rate
for the protocol when we do not have complete suppression of the
tunneling. For a device to successfully perform the protocol we require the
contrast ratio to be high, reaching a maximum of 1 when complete
suppression is possible. This can be determined by calculating the product of
the overlap $\langle 1 | \mathcal{D}_0\rangle$ at $t=0$ and $\langle
\mathcal{D}_0 | 5\rangle$ at t=$t_{\max}$.

\begin{equation}
 \langle 1 | \mathcal{D}_0 (t=0)\rangle =
  \frac{\Omega_{2\max}^2}{\sqrt{\Omega_{1\min}^4+\Omega_{2\max}^4 +
      \Omega_{1\min}^2\Omega_{2\max}^2}},
\end{equation}
and assuming $\Omega_{1\min} \ll \Omega_{2\max}$, to first order:

\begin{equation}
\langle 1 | \mathcal{D}_0 (t=0)\rangle =1-\frac{\Omega_{1\min}^2}{2\Omega_{2\max}^2}.
\end{equation}
Similarly, at the end of the protocol, when $t = t_{\max}$ the null
state is:

\begin{equation}
 \langle \mathcal{D}_0 (t=t_{\max})| 5 \rangle =
  \frac{\Omega_{1\max}^2}{\sqrt{\Omega_{2\min}^4+\Omega_{1\max}^4 +
      \Omega_{2\min}^2\Omega_{1\max}^2}}.
\end{equation}
Applying the same approximations as before gives the result:
\begin{equation}
\langle \mathcal{D}_0 (t=t_{\max})| 5 \rangle = 1-\frac{\Omega_{2\min}^2}{2\Omega_{1\max}^2},
\end{equation}
so a first approximation to the fidelity of population transfer in
this case of imperfect contrast ratio is

\begin{equation}
\left | \langle 5 | \mathcal{D}_0 (t=t_{\max})
\rangle\langle\mathcal{D}_0 (t=0) | 1 \rangle \right |^2 = 1 -
\frac{\Omega_{1\min}^2\Omega_{2\min}^2}{8\Omega_{1\max}^2\Omega_{2\max}^2}.
\label{eq:contrastratio}
\end{equation}
To illustrate with an example, to achieve a fidelity of 99.9\% and
assuming that $\Omega_1=\Omega_2$, we require a ratio of
$\Omega_{\min}/\Omega_{\max}=0.3$, hence we can see that a high
fidelity transfer may still be possible, even with imperfect controls.

We now turn our attention to calculating the gate modulated tunnel matrix elements
for a semi-realistic 5 donor device which we will use with the above
expressions for adiabaticity and fidelity to estimate the timescale
of the protocol over such a device.

\section{Numerical Modeling}
\label{sec:nummodel}

\begin{figure}
\begin{centering}
\includegraphics[height=5cm]{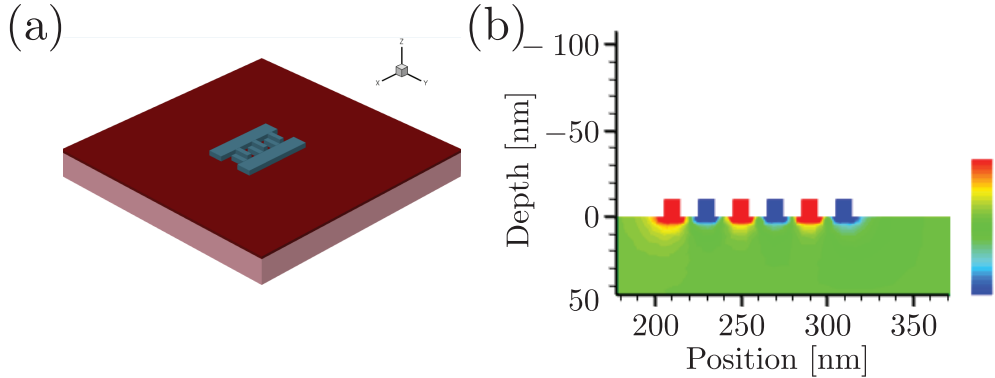}
\caption{a) An interdigitated gate design for A-CTAP. Donors are placed in between gate
  ``fingers'' in the central region of the device. All odd gates are
  connected together and similarly for all the even gates. b) A slice
  taken from a TCAD simulation of potentials within the device,
  slicing through the centre of all the donors which are located in a
  line parallel to the position axis. \label{fig:interdigit}}
\end{centering}
\end{figure}

The increased simplicity of an alternating coupling scheme over other
schemes makes it an attractive candidate for possible implementation
for demonstrating CTAP in a Si:P setting. 
Figure \ref{fig:interdigit} illustrates a possible gate design to implement
ACTAP$_5$. This design features global control of
the even and odd tunneling matrix elements with even and
odd barrier gates connected to be pulsed together. These global
controls are desirable due to the simpler fabrication and reduced
control circuitry required. Fine tuning of the end of chain sites may
be implemented using further gates but they are not included
here. Modeling of the potentials within the device with applied gate
biases was performed using a commercial TCAD finite element modeling
package. A simulation of the correct gate voltage sequence required to
enact the CTAP protocol is a difficult control problem with cross-talk
effects needing to be taken into account. Compensation schemes for
gated donor schemes have been investigated in Ref.
\cite{bib:KandasamyNano2006} and similar measures may need to be
implemented here to ensure the correct tunnel matrix elements are
obtained. We apply here a simpler approach of looking at pairwise tunneling
rates with a simplified structure consisting of two donors and a
barrier gate to modulate tunneling rates.

\begin{figure}
\begin{centering}
\includegraphics[height=5cm]{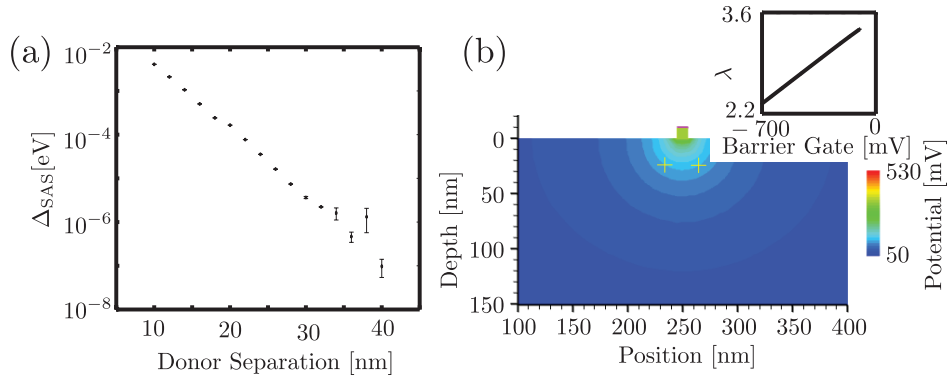}
\caption{(a) Ungated $\Delta_{SAS}$ as a function of donor separation in
  [100]. As donors are moved further apart the tunneling rate between them
  is decreased. The large fluctations at large donor separations are due to
  fluctations in the Monte Carlo routine. (b) A slice taken from a TCAD simulation of the simplified,
  one gate, two donor device. The slice in the vertical plane of the
  two donors. The $10~\textrm{nm}$ wide gate is centred about
  $250~\textrm{nm}$ directly between the positions of the two donors placed $30~\textrm{nm}$ apart, as marked by the crosses in the picture. (Inset) Relationship between fitted
  charge density and TCAD  obtained fields. Note the linear
  fit. Errors in the value of $\lambda$ obtained were $\pm$0.01\%.
\label{fig:dsasvspacing}}
\end{centering}
\end{figure}

The phosphorus in silicon scheme we are considering consists of
ionised $\mathrm{^{31}P}$ buried in a $\mathrm{^{28}Si}$ substrate. Each
of the ionised phosphorus atoms forms bonds with neighbouring silicon
atoms with their remaining 4 valence electrons. One un-ionised donor remains in
the chain with a remaining valence electron loosely bound to the donor. This
remaining electron's wavefunction may be manipulated using surface
gates to effect the transport across the chain.

The electronic states of shallow donors in silicon have been well
studied. To calculate the energy states of the remaining valence
electron we use the effective mass approach developed by Kohn and
Luttinger \cite{bib:KohnLuttinger1955} in which the donor electron is
described using hydrogenic envelope functions. This approach is
extended to include a second ionised donor such that the electron is
described by symmetric and anti-symmetric superpositions of the ground
state for each donor and an external electric field. The
effective Hamiltonian for the two donor, Si:P-P$^+$ system is
given by:  

\begin{equation}
\mathcal{H} = \mathcal{H}_{\mathrm{si}} + V_d(r-R_1) + V_d(r-R_2) + V_E
\end{equation} 
where $H_{\mathrm{Si}}$ is the Hamiltonian of an electron in the pure silicon lattice, which includes both a kinetic term and the
effective potential due to the silicon lattice,  $V_d$ the
coulombic potentials of each donor and $V_E$ an external applied field
\cite{bib:KoillerPRB2006, bib:WellardPRB2006, bib:ConradPhD2007}. 
The valence electron can be described by symmetric and anti-symmetric
superpositions of the ground state of each for each separate donor:
\begin{equation}
\Psi_\pm(\mathbf{r})=\mathcal{N}[\psi(\mathbf{r}-\mathbf{R}_\alpha)\pm\psi(\mathbf{r}-\mathbf{R}_\beta)]
\end{equation}
where $\mathcal{N}$ is a normalisation factor and the the ground state for a single phosphorus donor centred about
$R_\alpha$ is given by:

\begin{equation}
\psi(r)=\sum^6_{\mu=1}F_\mu(r-R)e^{\imath k_\mu\cdot(r-R_\alpha)}.
\end{equation}
The $F_\mu(r-R)$ are effective mass, hydrogenic envelope functions about the six
degenerate band minima.

TMEs may then be calculated by examining the minimum
energy splitting between the lowest energy states of a donor pair, the
symmetric-antisymmetric gap $\Delta_{\mathrm{SAS}}$. TMEs may
then be varied from $\Omega_{\max}=\Delta_{\mathrm{SAS}}$ to 0 using
time varying gate bias pulses such as those described in equations
\ref{eq:tmes} and \ref{eq:tmes2}. Examining these gate driven couplings gives an indication
of the timescale of the tunneling rates, and hence the timescale of
the protocol. 

Figure \ref{fig:dsasvspacing} shows the calculated $\Delta_{\mathrm{SAS}}$ as a
function of donor spacing for a P-P$^+$ system with no external field is
applied. At 30nm  $\Delta_{\mathrm{SAS}}$ is calculated as $0.1\mathrm{\mu
  eV}$, corresponding to a tunneling matrix element of $\sim 2~\textrm{GHz}$. These
calculations are consistent with the hydrogenic results of Openov \cite{bib:OpenovPRB2004} 
To include the effect of the external field on the tunneling rate we
have utilised an analytic potential fitted to the potential landscapes obtained
from several TCAD simulations at different gate biases. This approach
allows us to use a simple form for the external field which varies
smoothly over space which can be varied with any possible gate potential while
maintaining a close connection to modeling of semi-realistic
structures. In the y-z plane we approximate the surface gate with an infinite line of
charge the same distance above the ground plane with the origin at the
line of charge. Using the method
of images we obtain the form of the potential:

\begin{equation}
V = \frac{\lambda}{2 \pi \epsilon_0\epsilon_r}\ln{\frac{y^2+(z+d)^2}{y^2+(z-d)^2}}
\end{equation}
where $\lambda$ is the charge density of the line of charge obtained using the
simulated TCAD potentials, shown in figure \ref{fig:dsasvspacing}(b)Inset and d is the distance of the line of charge
above the ground plane. Since the dielectric constant of the SiO$_2$ is
much less than that of the bulk silicon, the discontinuity in the
resulting field is accounted for by extending the 5nm oxide layer to
an equivalent thickness with a bulk silicon dielectric constant to ensure the
correct field in the bulk silicon is obtained. This relationship between $\lambda$ and
barrier gate voltage has a constant offset due to the contact
potential of around $0.5~\textrm{V}$
that arises from the aluminium gates placed on the silicon dioxide
layer. Since we are only interested in the effect net effect of the gate 
we have ommitted this constant offset in the calculation of
$\Delta_{\mathrm{SAS}}$ so zero volts corresponds to the field-free
(ie purely hydrogenic) case. In practice this would be achieved by the
gate offsetting the surface charge.

\begin{figure}
\begin{centering}
\includegraphics[height=5cm]{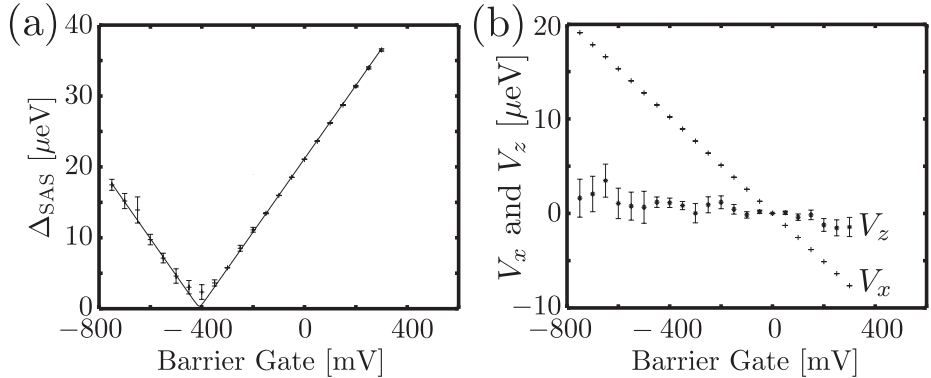}
\caption{(a)Variation of $\Delta_{\mathrm{sas}}$ with applied barrier gate
  voltage. We see a variation in $\Delta_{\mathrm{sas}}$ from $0.5$ to $10~\textrm{GHz}$ in a
  change of approximately $250~\textrm{mV}$ applied voltage on the barrier
  gate. The larger errors at larger spacings are due to the Monte
  Carlo routine. (b) $\mathrm{v_x}$ and $\mathrm{v_z}$
  coefficients as a function of barrier gate voltage. The fluctations
  in $\mathrm{v_z}$ are due to sampling errors in the Monte Carlo routine.\label{fig:dsasplot}}
\end{centering}
\end{figure}

Figure \ref{fig:dsasplot} shows the calculated variation of
$\Delta_{\mathrm{SAS}}$ with barrier gate voltage. We see a variation in
$\Delta_{\mathrm{SAS}}$ from $\sim 0.5$ to $10~\textrm{GHz}$ with a
  change of approximately $250~\textrm{mV}$ applied voltage on the barrier gate.  This is
  consistent, within uncertainty, with previous modelling of the energy gap for P-P$^+$ charge qubit with donor
  spacing of 30nm, showing that tunneling may be varied from around 0 to $10~\textrm{GHz}$
  with a change in barrier gate bias of $300~\textrm{mV}$
  \cite{bib:HollenbergPRB2006}. The errors in $\Delta_{\mathrm{SAS}}$ are
  dominated by fluctuations in $\mathrm{v_z}$ due to sampling errors in
  the Monte Carlo routine. Using equation \ref{eq:contrastratio} with 
  $\Omega_{\min}=0.5$GHz and $\Omega_{\max}=10$GHz
  and assuming $\Omega_1 = \Omega_2$ we obtain an error rate of
  $10^{-6}$. This result demonstrates the remarkable robustness of the
  protocol to imperfect control of the tunneling matrix elements. Ideally $\mathrm{v_z}$ is 0 when the field
  is completely symmetric with respect to the two donors and consists
  of the diagonal elements of the external field component of the
  Hamiltonian cancelling with each other. If we neglect $\mathrm{v_z}$
  in calculating $\Delta_{\mathrm{SAS}}$ we obtain a minimum tunneling
  rate of around 0.1GHz and an error rate of $10^{-8}$.

Assuming perfect contrast, for a target adiabaticity of 0.01 we may now calculate the required time
for the protocol. 
Using equation \ref{eq:adiabatsimpagain}, with
$\Omega_{\max}$=10GHz we obtain $t_{\max}\sim 70$ns as the required time
for adiabatic transfer of an electron across the chain, in this case
120nm long. 

Allowing for a large range variablity in the TMEs while still being
able to obtain high fidelity transfer of the electron means that the
protocol is also robust to donor placement. Variations to donor
placement will result in variations in the ungated TMEs as shown in
Figure \ref{fig:dsasvspacing}. However, as shown in equation
\ref{eq:contrastratio}, adiabatic behaviour will still occur over a
wide range of maximum and minimum TMEs. As a result the protocol can
be insensitive to variations in donor placement, which is important
since devices such as this proposed are generally fabrication using
ion implantation techniques where there is variability in the exact
placement of the donor.

\section{Conclusion}
We have shown that for a semi-realistic device consisting of buried
phosphorus donors in silicon with surface gates, an adiabatic pathway
may be found allowing transport of quantum information along a donor
chain using the alternating coupling scheme. The fidelity of the transport protocol is resilient to variations in the
TMEs along the chain so long as the end-of-chain sites can be
controlled. We calculate that, for a device with a relatively large donor spacing of 30nm, with a maximum pairwise tunneling rate of 10 GHz which can be suppressed to 0, we can estimate a time of $\sim70$ ns required for adiabatic operation. For closer positioned donors we expect an adiabatic operation time down to the few nanosecond regime or faster. This time is well within measured electron spin relaxation times
and dephasing for silicon and within the expected charge
relaxation times for Si:P devices.

\ack
This work was supported by the Australian Research Council Centre of
Excellence Scheme, the Australian Government and by the US National Security Agency (NSA), Advanced Research and Development Activity (ARDA) and the Army Research Office (ARO) under Contract No. W911NF-04-1-0527.  ADG is the recipient of an Australian Research Council Queen Elizabeth II Fellowship (project number DP0880466), LCLH is the recipient of an Australian Research Council Australian Professorial Fellowship (project number DP0770715).


\end{document}